\documentclass[%
 reprint,
 preprintnumbers,
 nofootinbib,
 amsmath,amssymb,
 prd,
]{revtex4-2}

\usepackage{graphicx}
\usepackage{dcolumn}
\usepackage{bm}
\usepackage{booktabs}
\usepackage{xcolor}
\usepackage{slashed}
\usepackage{amssymb}
\usepackage{lipsum}
\usepackage{multirow}
\usepackage{url}
\usepackage[normalem]{ulem}
\usepackage[hidelinks]{hyperref}

\begin{document}

\preprint{UCI-TR-2025-25}

\title{Statistics meet systematics: Resolution of the massive early JWST galaxy tension}

\author{Jay R.\ Krishnan}
\email{radhakrj@uci.edu}
\affiliation{Center for Cosmology, Department of Physics and Astronomy,  University of California---Irvine, Irvine, California 92697-4575, USA}

\author{Kevork N.\ Abazajian}
\email{kevork@uci.edu}
\affiliation{Center for Cosmology, Department of Physics and Astronomy,  University of California---Irvine, Irvine, California 92697-4575, USA}

\date{\today}

\begin{abstract}
The discovery of massive, high redshift galaxies with the James Webb Space Telescope (JWST) has been argued to challenge $\Lambda$CDM (cold dark matter): such systems would require extremely rare halos and baryon-to-stellar-mass conversion efficiencies unphysically approaching—or exceeding—$100\%$. If confirmed at galaxy-formation--forbidden efficiencies, these galaxies could signal new physics beyond standard cosmological structure formation. We develop a galaxy model framework that ties the linear power spectrum to the inferred efficiencies of galaxy growth while incorporating multiple sources of uncertainties in order to test the structure formation models. The sources of error include (i) observational sample variance, (ii) asymmetric scatter induced by the steepness of the high-mass halo tail, and (iii) systematic uncertainties in stellar mass estimates. We find that the inferred star-formation efficiency is largely controlled by systematic uncertainties in the stellar mass estimates derived from spectral energy distribution modeling of JWST-detected galaxies. Because of the inherent Eddington-like bias, systematic uncertainties amplify the asymmetry of the scatter, in some cases by orders of magnitude, thereby bringing the inferred efficiencies into closer agreement with expectations from early galaxy formation models. We also present how these uncertainties can be applied to the inferred UV luminosity function. Our framework can be used to test $\Lambda$CDM as errors are reduced and further detections are made.

\end{abstract}

\maketitle


\section{\label{sec:intro}Introduction}

Since its launch, the James Webb Space Telescope (JWST) has uncovered numerous high-redshift ($z$) galaxy candidates, several of which appear anomalously luminous and massive for their observed epochs. A rapidly growing literature has examined the implications of these objects~\cite{Boylan-Kolchin:2022kae, Yung:2023bng, Yung:2025ttv, Shen:2024hpx, chworowsky_evidence_2023}. In particular, Boylan-Kolchin (BK) ~\cite{Boylan-Kolchin:2022kae} showed that the abundance of sufficiently massive dark-matter halos at $z\gtrsim 8$ is inadequate to host the reported stellar masses when compared via cumulative comoving stellar-mass density. 

Recent work looking at large samples of galaxies at $z>4$ in the Cosmic Evolution Early Release Science (CEERS) Survey infers that the efficiencies of conversion of baryons to stars increases with redshift, and that these efficiencies may be consistent with that expected in high-$z$ galaxy and star formation \cite{chworowsky_evidence_2023}. It has been argued that at $z > 10$, global baryon conversion efficiency $\epsilon$ may approach unity due to  a high interstellar medium density in this epoch, along with low metallicities, which allows molecular cloud free fall into star formation to be fast relative to the time scale for stars to develop winds and supernovae, which quench star formation in the local Universe \cite{Dekel:2023ddd}. The rapid assembly history of galaxies in this epoch is also seen in extrapolations to high resolution, high-$z$ simulations' results \cite{Keller:2022mnb,McCaffrey_NoTension_2023} and in semianalytic models \cite{Robertson:2022gdk}. Our goal is to build upon these prior studies by establishing a direct connection between the statistical properties of the observations and the underlying large-scale structure (LSS) and halo populations. 

One aim of our work is to provide a robust statistical test that incorporates random and systematic uncertainties in a framework that can be used to test for the possibility of new physics as responsible for the JWST high-$z$ galaxies, if studies find them to be discrepant with standard structure formation. Several proposals exist for new physics that can enhance early galaxy formation via augmenting the halo mass function (HMF) at its high-mass exponential cutoff \cite{CosmoVerseNetwork:2025alb}, including: dynamical dark energy \cite{Menci:2022wia} and enhanced clustering that may be present with the formation of massive primordial black holes \cite{Liu:2022bvr}; non-Gaussianities in the initial perturbation spectrum can also enhance the HMF at these early high-mass scales \cite{Biagetti:2022ode}; properties of the dark matter, such as augmentation of clustering via miniclusters formed from axions may boost early galaxy formation \cite{Hutsi:2022fzw}, axion quark nuggets \cite{Zhitnitsky:2023znn}, domain walls from axionlike particles \cite{Guo:2023hyp}, or fragmentation in ultralight axion dark matter \cite{Bird:2023pkr}, as well as the possibility of a negative cosmological constant in this epoch \cite{Adil:2023ara}.

In this work, we revisit the abundance-matching method of BK with new sources of uncertainty explicitly included. Specifically, we include sample variance on the galaxy counts, and systematic uncertainties present from spectral energy distribution (SED) modeling of the host galaxy. We also explore uncertainties associated with the choice of adoption of the HMF. For the first time, we show that the systematic error is significant for these galaxies as it propagates through the stellar mass density inferred from the galaxy to the efficiency of star formation. Namely, due to the rapidly falling nature of the HMF for these highest-mass objects, the upscatter of lower-mass galaxies dominates over the scatter in the downward direction \cite{Lima:2005tt}. This Eddington-like bias alters the inferred efficiency of star formation and degree of rarity for these detections. This bias introduced by the uncertainties in luminosity and magnitude \cite{2005MNRAS.362..321B,Teerikorpi:2015pda}, and the importance of their combination with extreme value statistics, has been previously considered \cite{Lovell:2022bhx}, as well as the effects of upscatter dominance for low-$z$ cluster surveys \cite{Lima:2005tt}. However, the effects on the high-$z$ stellar mass density due to the steeply falling underlying dark matter HMF at high masses have not been considered in combination with systematic uncertainties. Our work shows that this Eddington-like bias asymmetry in scatter, in combination with systematic errors in this context, should not be ignored.  

In our analysis, we use a broad sample of high-$z$ JWST galaxies, including early photometric candidates~\cite{Labbe:2022ahb}, spectroscopically confirmed systems~\cite{haro_spectroscopic_2023, Bunker:2023lzn, castellano_jwst_2024, carniani_spectroscopic_2024, Naidu:2025xfo}, and more recent identifications~\cite{Perez-Gonzalez:2025bqr}, shown in Table~\ref{tab:gal data}. In Sec.~\ref{sec:HMF} we examine the HMF, the modeling uncertainty it introduces, and construct the cumulative stellar mass density to test consistency with $\Lambda$CDM (cold dark matter). We then incorporate (i) sample variance from large-scale structure, (ii) asymmetric scatter arising from the steep high-mass tail of the HMF, and (iii) systematic uncertainties in SED-derived stellar masses in Secs.~\ref{sec:samplevar}–\ref{sec:syserror}. Results are presented in Sec.~\ref{sec:results}. We discuss methods of inclusion of statistical and systematic uncertainties in theory and measurement of the UV luminosity function, a complementary method often used to test galaxy formation, in Sec.~\ref{sec:lum}. We present our conclusions in Sec.~\ref{sec:conc}. Our analysis shows that the inferred baryon-to-stellar-mass conversion efficiency is dominated by systematic uncertainties in SED-based stellar masses and their amplification of Eddington-like bias. Accounting for these effects substantially alleviates the apparent tension between the most massive JWST galaxies and $\Lambda$CDM abundance expectations. \vskip -0.5cm


\section{\label{sec:methods}Theory and Observational Error}

Here, we explore the model uncertainties, observational uncertainties, and how those two combine in inferring a tension, or not, from the observed high $z$ galaxies. We first look at the model dependencies associated with adoption of the underlying HMF, which we find important, but not significant for HMFs calibrated for high-$z$, and high halo masses relevant for these galaxies. Second, we look at the underlying sample variance present due to underlying LSS for the observational samples, which is not uniformly included in reported observational data. We find that sample variance is nontrivial, but is not the largest effect. Lastly, we look at the Eddington-like bias introduced by both the random and systematic uncertainties due to the underlying steeply falling HMF and related stellar density function. This last combination of effects provides the largest source of alleviation of tension between the inferred and observed stellar mass densities at high $z$. \vskip -0.5cm

\subsection{\label{sec:HMF}Halo mass function and stellar mass density}

\begin{figure}[t!]
\centering
\includegraphics[width=0.48\textwidth]{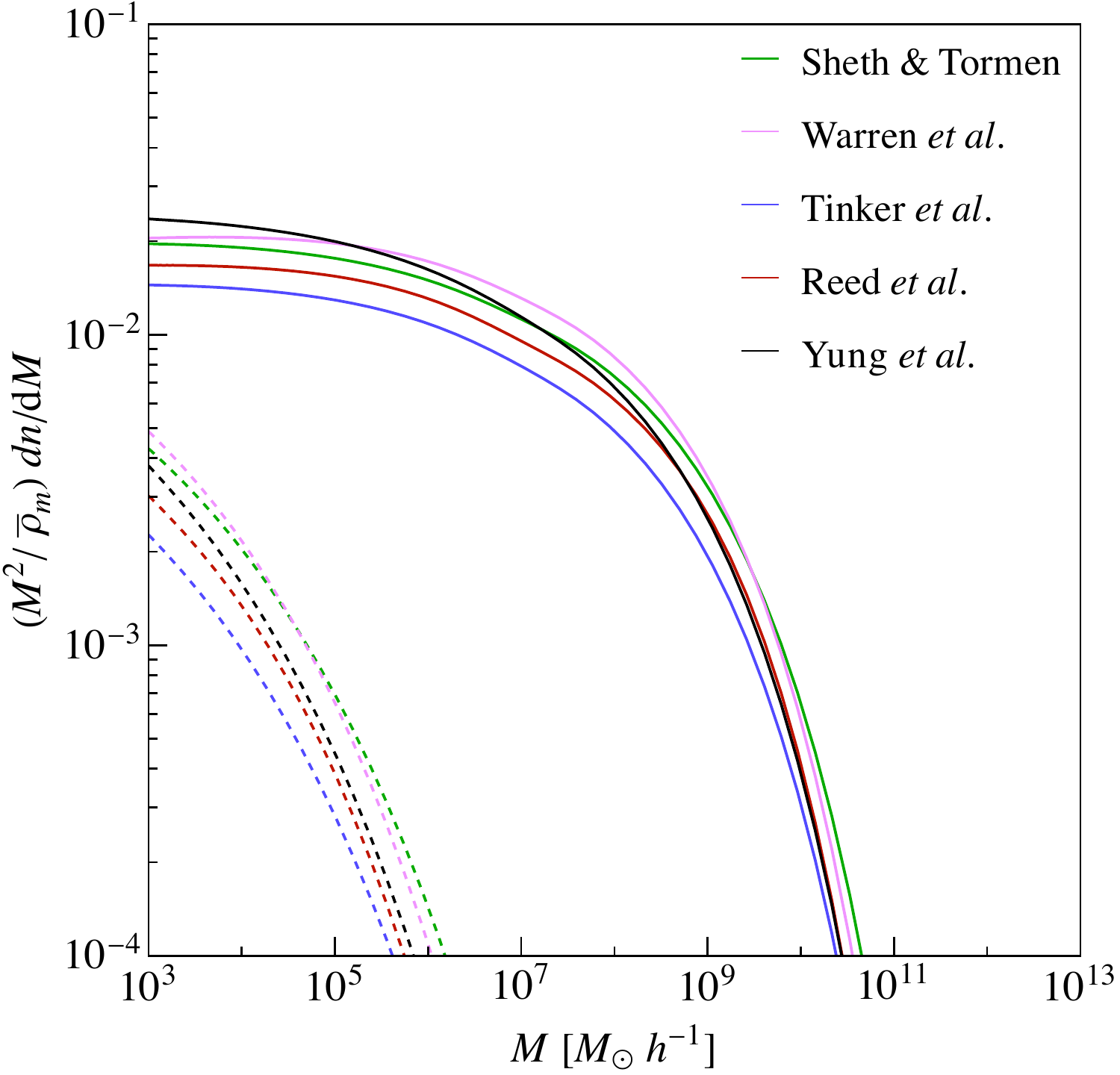}
\caption{Comparison of halo mass functions at redshifts 10 (solid lines) and 25 (dashed lines)~\cite{Sheth:1999mn, Warren:2005ey, Tinker:2008ff, Reed:2006rw, Yung:2025ttv}. Among these, the Reed and Yung HMFs are calibrated for use at high $z$. There are clear significant differences of the HMFs at smaller masses for $z=10$.  However, Reed \textit{et al.}\ and Yung \textit{et al.}\ are not very discrepant at the high-$M$ tail, especially when converting to cumulative comoving stellar mass density (see Fig.\ \ref{fig:MBK Reed vs Yung and sample var}). \label{fig:HMF comparison}
}\vskip -0.5cm
\end{figure}

Galaxies are hosted in dark matter halos that form from gravitational collapse of an initially linearly growing perturbation distribution (for a review, see Ref.~\cite{Cooray:2002dia}). We calculate the linear perturbation power spectrum using \texttt{CLASS} \cite{Blas:2011rf} for a flat, cosmological-constant, cold dark matter cosmology ($\Lambda$CDM), inferred from the measurements of the \textit{Planck} observatory: $H_0=67.32~\mathrm{km\,s^{-1}Mpc^{-1}}$, $\Omega_b = 0.0493$, $\Omega_m = 0.3158$, $\sigma_8 = 0.8120$, and $n_s = 0.96605$ (Table 1 from \cite{Planck:2018vyg}). Calculations of all quantities described below are available via a \textit{Mathematica} notebook\footnote{\url{https://github.com/jaykrishnan9121/JWST_Early_Galaxies}} \cite{JWSTEG}. The root-mean-square variance in the linear density field smoothed with a top-hat filter of radius $R = (3M/4\pi \bar{\rho}_m)^{1/3}$ is then \cite{Cooray:2002dia},
\begin{equation}
\sigma^2(M,z) = \dfrac{1}{2 \pi^2}\int dk\,k^2\, P(k,z)\,|W(kR)|^2\, ,
\end{equation}
where $P(k,z)$ is the linear matter power spectrum, and $W(kR)$ is the Fourier transform of the top-hat window function defined as~\cite{Cooray:2002dia},
\begin{equation}
W(kR) = \dfrac{3}{(kR)^3} \left[\sin(kR) - kR \cos(kR)\right]\, .
\end{equation}
The HMF is the number of dark matter halos with mass $M$ at redshift $z$ per unit mass per unit comoving volume \cite{Tinker:2008ff},
\begin{equation}
\dfrac{dn(z,M)}{dM} = f(\sigma) \dfrac{\bar{\rho}_m}{M}\dfrac{d\ln\sigma^{-1}}{dM}\, ,
\end{equation}
where 
\begin{equation}
f(\sigma) = A \left[\left( \dfrac{\sigma}{b}\right)^{-a} + 1 \right] e^{-c/\sigma^2}\, .
\end{equation}
The parameters in $f(\sigma)$ are acquired through $N$-body simulations. These are calibrated for different $z$ ranges and cosmologies and will predict different halo abundances. In principle, the level of anomaly of the inferred density of galaxies of a certain mass depends on the adopted HMF.  In Fig.\ \ref{fig:HMF comparison}, we show the HMFs from Sheth and Tormen (ST)~\cite{Sheth:1999mn}, Warren \textit{et al.}~\cite{Warren:2005ey}, Tinker \textit{et al.}~\cite{Tinker:2008ff}, and Reed \textit{et al.}~\cite{Reed:2006rw} as well as the one calculated in Yung \textit{et al.}~\cite{Yung:2025ttv}. 

The HMF from ST is widely used, while Warren \textit{et al.}\ provided one of the first high-resolution high dynamic mass-range $N$-body based fits. Tinker \textit{et al.}\ extended these results to better capture cosmology dependence and it is frequently adopted by low-$z$ galaxy cluster surveys. Reed \textit{et al.}\ focus on high-$z$ halos, and Yung \textit{et al.}\ provide a recent high-$z$ calibration consistent with recent cosmological parameters. Figure \ref{fig:HMF comparison} shows a comparison of these HMFs at redshifts 10 and 25. There are significant differences of the HMFs, particularly at smaller masses for $z=10$. However, the two high-$z$ calibrated mass functions, Reed \textit{et al.}\ and Yung \textit{et al.}\ are not highly discrepant at the high-mass tail, especially when converting to cumulative comoving stellar mass density, as we show in the next Section. 
 
\begin{figure}[t!]
\centering
\includegraphics[width=0.48\textwidth]{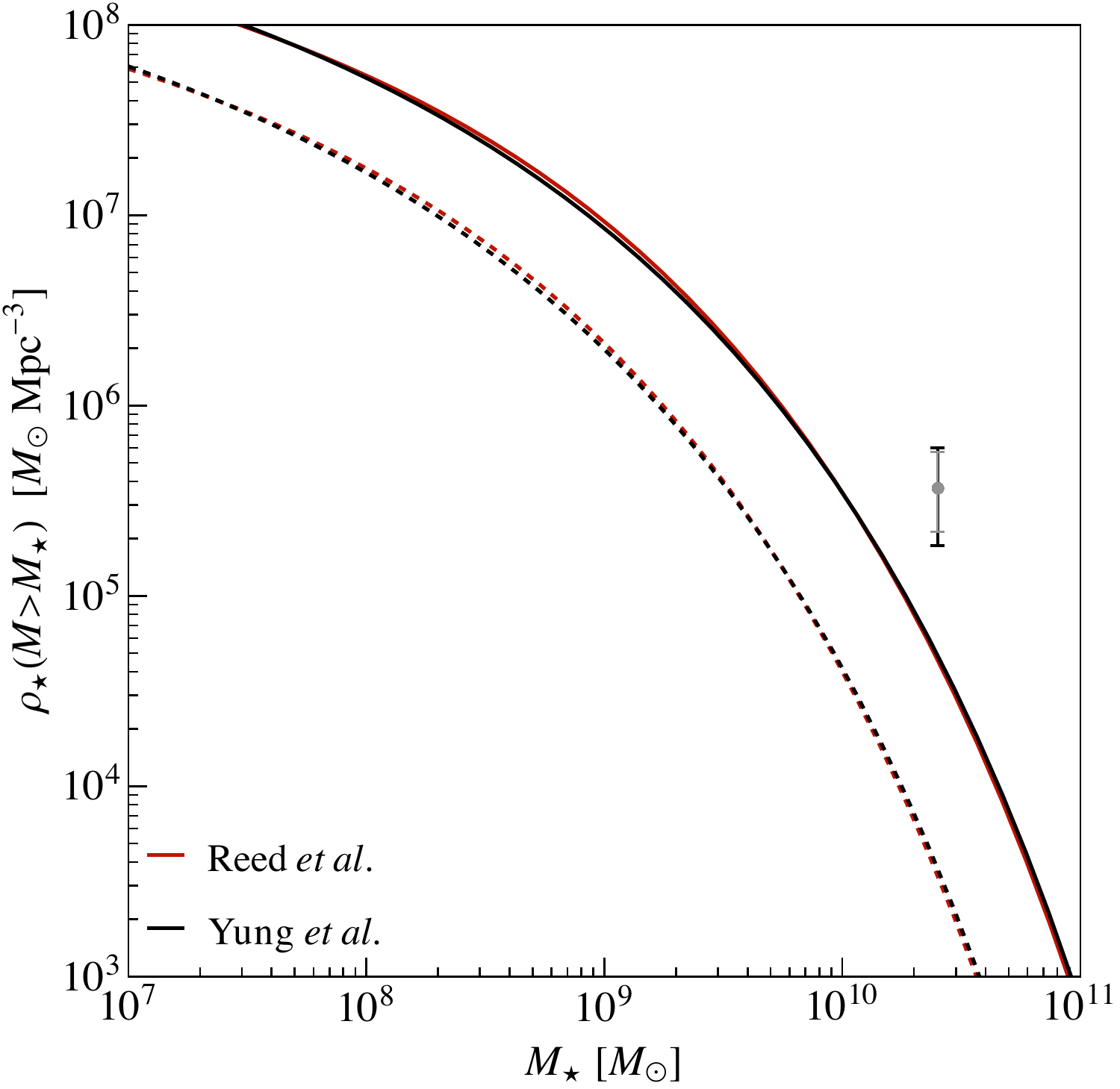}
\caption{Massive Labbé \textit{et al.}\ galaxy at $z = 9.08$ shown with cumulative comoving stellar-mass density curves computed using the Reed \textit{et al.}\ and Yung \textit{et al.}\ HMFs. Solid curves correspond to a baryon-to-stellar-mass conversion efficiency of $\epsilon = 1$, while dashed curves show $\epsilon = 0.5$. The figure implies that this galaxy's stars were formed at well over $100\%$ efficiency. Random errors for the galaxy are shown in gray, and random plus sample-variance errors in black. At the high-mass, high-$z$ regime probed by JWST, these HMFs yield only minor differences, and we adopt Yung \textit{et al.}\ for subsequent calculations. \label{fig:MBK Reed vs Yung and sample var}
}\vskip -0.5cm
\end{figure}

We assume that a dark matter halo of mass $M_{\rm halo}$ contains $f_bM_{\rm halo}$ in baryons, proportional to the critical density in baryons relative to matter, $f_b\equiv \Omega_b/\Omega_m$. If a fraction $\epsilon$ of those baryons is converted into stars, then a galaxy of stellar mass $M_\star$ must reside in a halo with $M_{\rm halo}\geq M_\star/(\epsilon f_b)$, a technique called abundance matching~\cite{Boylan-Kolchin:2022kae}. The cumulative comoving number density of halos more massive than $M_{\rm halo}$ at redshift $z$ is
\begin{align}
n(>M_\text{halo},z) = \int_{M_\text{halo}}^{\infty} dM \dfrac{dn(z,M)}{dM}\, .
\end{align}
Analogously, the cumulative comoving stellar-mass density in halos above $M_\star/(\epsilon f_b)$ is
\begin{align}
\rho_\star(>M_\star,z) = \epsilon f_b \int_\frac{M_\star}{\epsilon f_b}^{\infty} dM M \dfrac{d\bar{n}(z,M)}{dM}\, \label{eq:rhostar},
\end{align}
where the average of the HMF, $d\bar n/dM$, is over the survey volume to match the observational binning in $z$ and solid angle $\Omega_\text{FOV}$,
\begin{align}
\dfrac{d\bar{n}(z,M)}{dM} &= \dfrac{\Omega_\text{FOV}}{V_\text{bin}(\Omega_\text{FOV},z_1,z_2)}\int_{z_1}^{z_2}dz \,M \dfrac{dn(z,M)}{dM} \times \nonumber\\
&\qquad\qquad\qquad\dfrac{\chi(z)^2}{\sqrt{\Omega_m(1+z)^3 + (1-\Omega_m)}}\, .\label{eq:volAvgHMF}
\end{align}
This ensures a consistent comparison to the observed stellar mass density estimated as $M_\star$ divided by the comoving volume of the unit-$\Delta z$ bin and field of view that yielded the detection~\cite{Conselice:2024yls}. Due to unspecified survey volumes in some galaxy detections, we adopt the full volume of space in that JWST survey field used by the respective authors to find these massive high-$z$ galaxies, binned in unit $z$. This may be a conservative overestimate of the volume in the cases where this was applied (see Table~\ref{tab:gal data}). As illustrated in Fig.\ \ref{fig:MBK Reed vs Yung and sample var} for a representative galaxy from Labbé \textit{et al.}~\cite{Labbe:2022ahb} at $z=9.08$, the choice among high-$z$ HMF calibrations induces only minor differences in $\rho_\star$, and therefore we adopt the HMF from Yung \textit{et al.}~\cite{Yung:2025ttv} henceforth.\vskip -0.2cm

\subsection{\label{sec:samplevar}Sample variance}

LSS induces field-to-field fluctuations in number counts beyond Poisson noise. We account for this sample variance through the square root of the variance term from linear bias. The galaxies in our sample are at a large comoving distance, $r_i$, from us and were found in FOVs (shown in Table~\ref{tab:gal data}) of radius $\Theta_s \lesssim 10^{-3}\, \mathrm{rad}$. This justifies the small angle approximation, and we use the following expression from Eq.~(7) in Hu and Kravtsov~\cite{Hu:2002we} for the sample variance:
\begin{align}
\dfrac{\langle n_i^2 \rangle - \bar{n}^2}{\bar{n}^2} &= \dfrac{ b_i^2}{\delta r_i r_i^2} \int \dfrac{d^2\ell}{(2 \pi)^2} \left( 2 \dfrac{J_1(\ell \Theta_s)}{\ell \Theta_s}\right)^2 P(\ell/r_i,z_i) \nonumber\\\nonumber\\
&= \dfrac{ b_i^2}{\delta r_i r_i^2} \int_0^\infty \dfrac{d\ell}{2 \pi}  \ell \left( 2 \dfrac{J_1(\ell \Theta_s)}{\ell \Theta_s}\right)^2 P(\ell/r_i,z_i),
\end{align}
where $\delta r_i$ is the comoving distance associated with our galaxy binning, and $\ell$ is the product of the comoving distance and wave number.

Here, $P(k,z)$ is the linear matter power spectrum, $J_1$ is the first order Bessel function of the first kind, and $b_i$ is the linear halo bias~\cite{Sheth:1999mn}. We then combine in quadrature these errors, scaled with observed stellar density, with the statistical errors of the observed stellar mass density of each galaxy. The impact on the inferred $\epsilon$ for our sample is shown in Fig.\ \ref{fig:MBK Reed vs Yung and sample var}, and is not as significant as the systematic errors we consider later. \vskip -0.2cm

\subsection{\label{sec:assyminscatter}Bias in random error from falling distributions}

Given an observable $x$ that has some true distribution function $v(x)$, it imparts a bias on the observed distribution function $u(x)$. This effect, called Eddington bias, and methods to correct for it were first discussed in \citet{1913MNRAS..73..359E}. Such bias can be understood to impart a multiplicative shift in the two distribution functions, 
\begin{equation}
u(x) = \exp\left(\frac{\sigma}{2}\frac{d^2}{dx^2}\right)v(x)\,.\label{eq:numbercounts}
\end{equation}
In the case of an exponentially falling true distribution function, $v(x)=e^{-kx}$, common in astronomy, the bias in the observable can be given as a shift in the observable $x$. Instead of correcting for the reported number counts in Eq.~\eqref{eq:numbercounts}, one can correct for the reported observable, which for our purposes is the object mass, $M$. For the case of a log-normal distribution in $\ln M$, with random error $\sigma_{\ln M}$, $\Delta x = \ln M_\mathrm{edd} - \ln M_\mathrm{obs}$ is the shift between the true and observed mass that reproduces the same multiplicative shift in Eq.~\eqref{eq:numbercounts}. This is
\begin{equation}
\ln M_\mathrm{edd} - \ln M_\mathrm{obs} = -\frac{1}{2}k \sigma_{\ln M}^2\, , \label{eq:lnM}
\end{equation}
and since $k>0$, the observed mass of an object is shifted to be greater than its true mass. \textit{This} is what is traditionally referred to as the Eddington bias. The effect of this bias is significant, and was studied along with important considerations of extreme value statistics of these galaxies in \citet{Lovell:2022bhx}.

Since the underlying HMF is not purely exponential, we must adopt a formalism more flexible than Eddington bias for exponential distributions. A more flexible methodology was used by Lima and Hu~\cite{Lima:2005tt} (LH). That work was primarily focused on low-$z$ cluster surveys. Their considerations are similar to Eddington's, but are not for the observable quantity's distribution, such as magnitude, luminosity, or inferred stellar mass, but rather
this bias within the underlying framework of the dark matter halo model. The HMF is steep and nearly exponential at large masses, leading to the need for an Eddington-like correction.  Following LH, we correct for this Eddington-like bias by convolving the volume averaged HMF with an error function,
\begin{align}
\rho_\star^\text{LH}(>M_\star,z) = \epsilon f_b \int_0^{\infty} dM \dfrac{d\bar{n}(z,M)}{d \ln M} \times \nonumber\\
 \dfrac{1}{2}\text{erfc}\left[ \dfrac{\ln(M_\star/\epsilon f_b) - \ln M - \ln M_\mathrm{sys}}{\sqrt{2 \sigma_{\ln M}^2}} \right]\, . \label{eq:rhostarLH}
\end{align}
For illustration, in Fig.\ \ref{fig:MBK sys} we set $\ln M_\mathrm{sys} = 0$ and take $\sigma_{\ln M}$ from the lower mass uncertainty of the source. The resulting $\rho_\star$ curves, shown in purple, shift upward compared to those in Fig.\ \ref{fig:MBK Reed vs Yung and sample var} (equivalently, the required $\epsilon$ shifts downward), reflecting the expected Eddington-like bias. While the shift is small, this equation provides a consistent framework for incorporating systematic uncertainties. We find that the asymmetry bias correction $\ln M_\mathrm{edd} - \ln M_\mathrm{obs}$ using LH is smaller by a factor of $\sim\! 3$, on average, than the Eddington bias correction $-(1/2)k \sigma_{\ln M}^2$  in $\ln M$  Eq.~\eqref{eq:lnM}.

\begin{figure}[t!]
\centering
\includegraphics[width=0.48\textwidth]{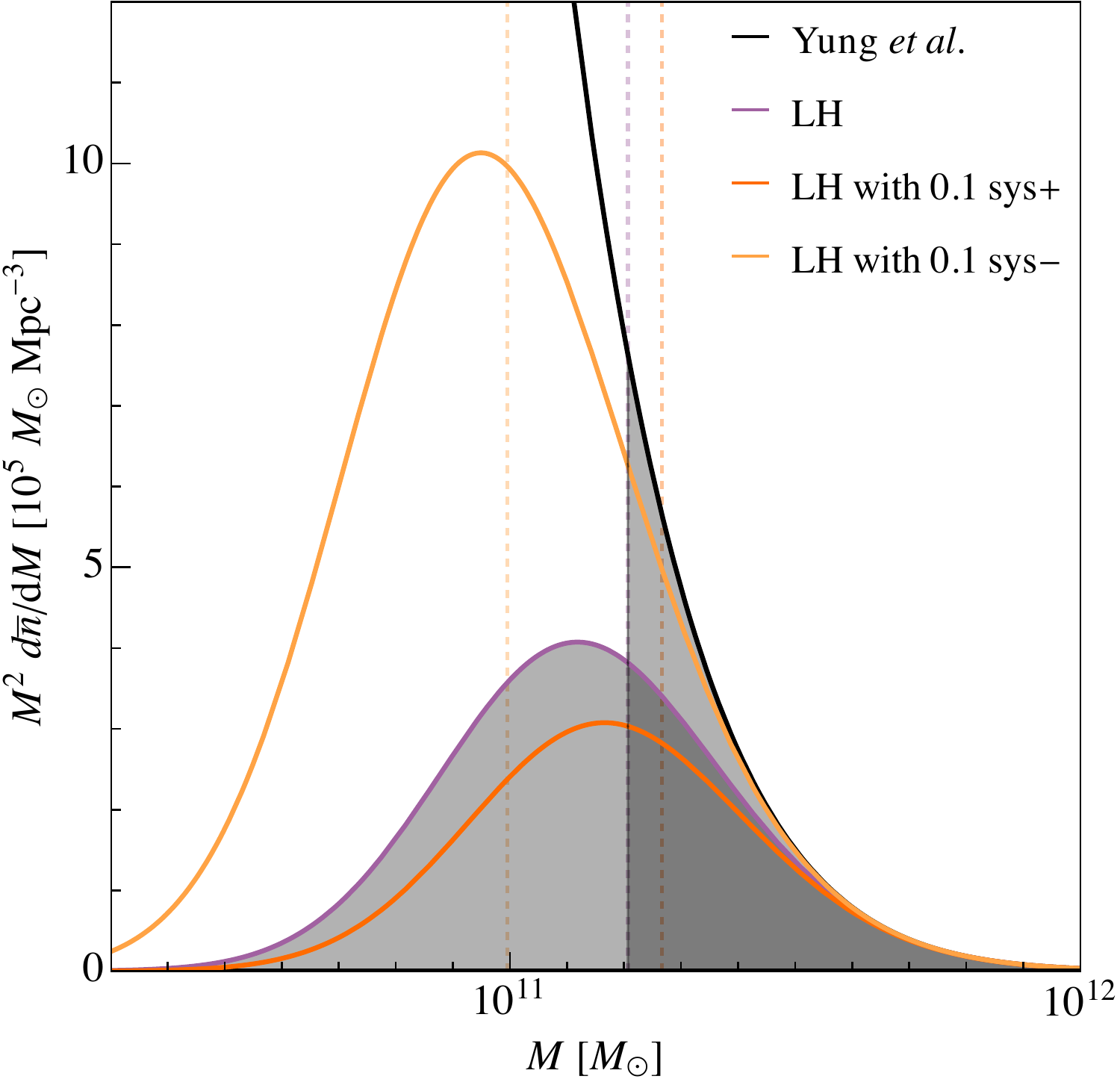}
\caption{Illustration of the scattering asymmetry in the cumulative halo mass density. The dashed purple line marks the halo mass corresponding to the Labbé \textit{et al.} galaxy at $z = 9.08$ assuming $\epsilon=1$. The black curve shows the Yung HMF; integrating it above $M_\text{halo}$ yields $\rho_\star$ [Eq.~\eqref{eq:rhostar}].
The purple curve shows the modified integrand in Eq.~\eqref{eq:rhostarLH}. The shaded regions represent halos scattering across $M_\text{halo}$: low-mass halos upscatter into the integral and high-mass halos downscatter out. In this case, upscattering dominates by $\sim\!4\times$, and produces a $80\%$ increase in $\rho_\star^\text{LH}$. This is an Eddington-like bias in the halo-model framework.
Systematic uncertainties shift the halo mass $M_\text{halo}$ to the dashed orange lines: dark orange for upward systematic errors and light orange for downward. The orange curves show the corresponding integrands (computed using only $10\%$ of the systematic uncertainties for clarity), illustrating that the downward shift yields a substantially larger change in the cumulative halo mass density than the upward shift.\label{fig:LH}}\vskip -0.5cm
\end{figure}

Figure~\ref{fig:LH} illustrates how we account for the scattering asymmetry, i.e., how Eq.~\eqref{eq:rhostarLH} leads to the upward shift we see in Fig.~\ref{fig:MBK sys}. The dashed vertical purple line is the halo mass corresponding to the massive Labbé \textit{et al.}\ galaxy at $z = 9.08$ assuming a conversion efficiency of $\epsilon = 1$ ($M_\text{halo}=M_\text{gal}/f_b$). The black curve shows the (nearly) exponential tail of the Yung HMF at $z = 9.08$ where this galaxy's halo resides. Calculating the integral in the cumulative comoving stellar mass density, Eq.~\eqref{eq:rhostar}, equates to computing the area under the black curve from $M_\text{halo}$ to infinity. The purple curve is the integrand in Eq.~\eqref{eq:rhostarLH}, which includes an error function with width that is controlled by the lower random error in the observed galaxy mass that. The error function smoothly cuts off the HMF at $M_\text{halo}$. The integrated region in Eq.~\eqref{eq:rhostarLH} now includes the light-gray region below the purple curve but excludes the light-gray region above it. These two regions above and below the integrand represent halos scattering across the integration boundary $(M = M_\text{halo})$: lower-mass halos ($< M_\text{halo}$) are upscattered into the integrated range, while higher-mass halos ($> M_\text{halo}$) are downscattered out of it. For the given case, the amount of upscatter is about four times larger than downscatter and leads to an $80\%$ increase for $\rho_\star^\text{LH}$ relative to $\rho_\star$ with an exact threshold. This is Eddington-like bias in the halo model framework, and results in the purple curve's upward shift in Fig.~\ref{fig:MBK sys} compared to what we see in Fig.~\ref{fig:MBK Reed vs Yung and sample var}.\vskip -0.2cm

\subsection{\label{sec:syserror}Systematic errors}

\begin{figure}[t!]
\centering
\includegraphics[width=0.48\textwidth]{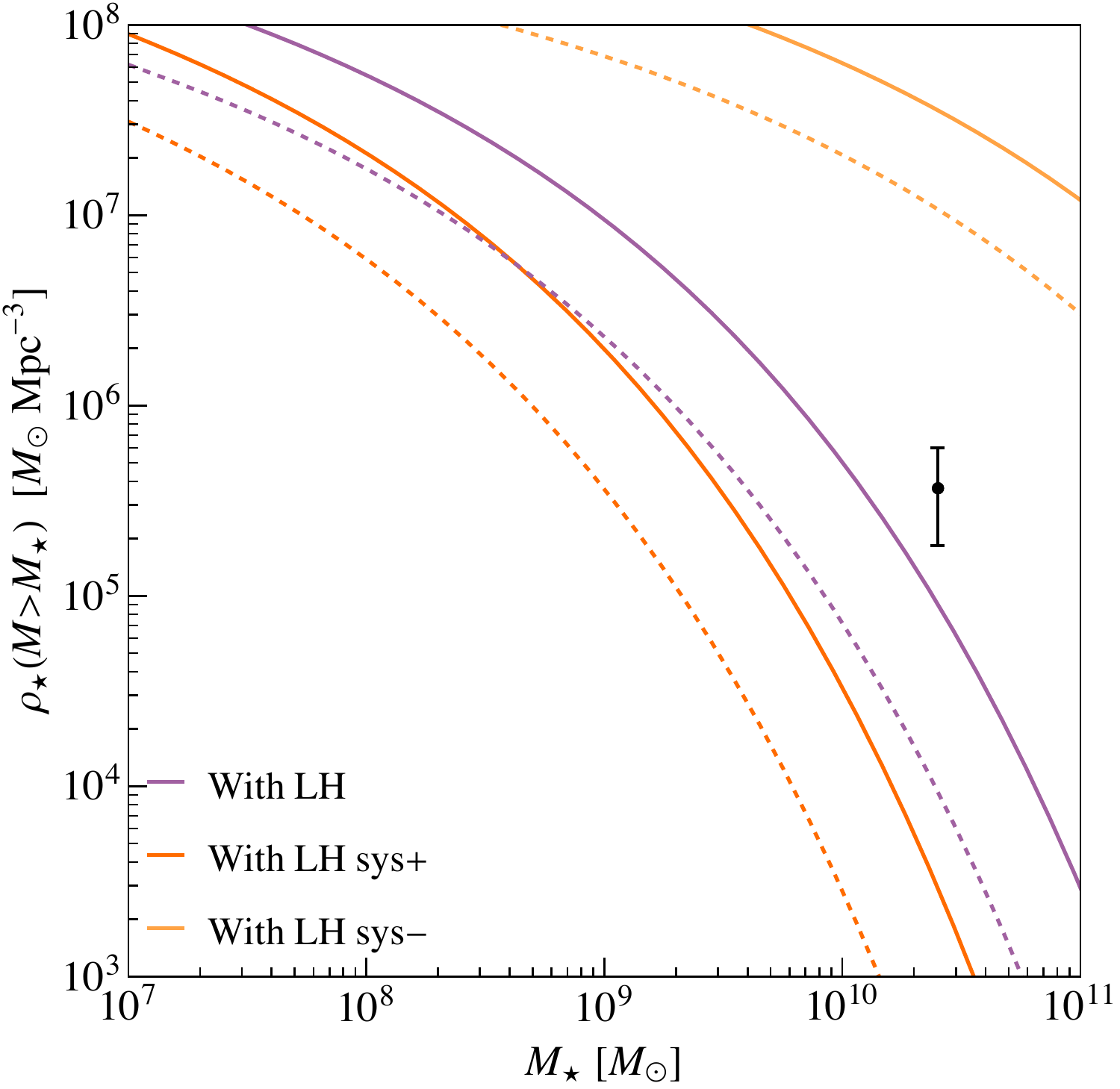}
\caption{Systematic uncertainties shown as two sets of curves. The purple stellar-mass density curves are computed using Eq.~\eqref{eq:rhostarLH} and shift upward relative to Fig.~\ref{fig:MBK Reed vs Yung and sample var} after we account for asymmetry in the random scatter. Error bars on the galaxy point represent random errors plus sample variance. The orange curves indicate the upper (dark orange) and lower (light orange) systematic bounds when inferring the efficiency associated with the galaxy. See text for further details. Comparing the galaxy’s central value with these curves shows that the total uncertainty is dominated by systematics. \label{fig:MBK sys}
}\vskip -0.5cm
\end{figure}

Stellar masses are derived from SED fitting and can vary substantially across modeling choices~\cite{Labbe:2022ahb, haro_spectroscopic_2023, Perez-Gonzalez:2025bqr}. We therefore treat SED-model dependence as a systematic uncertainty. Each SED derived galaxy mass value has a central value with upper and lower random errors. For galaxies where multiple SED pipelines are available, we take the median of central mass values as the galaxy's mass $m$. The same is done for the upper and lower random errors separately to get the final random errors $\delta m_\text{rand}^+$ and $\delta m_\text{rand}^-$. To bracket the systematic uncertainty, we take the largest deviation of the central mass value, among the different SED derived masses, from $m$ to get the upper and lower systematic errors $\delta m_\text{sys}^+$ and $\delta m_\text{sys}^-$.

When multiple SED pipelines are available, we adopt the median mass and take the extrema relative to the median as asymmetric systematic bounds. To be conservative, when only a single pipeline is available, as marked in Table~\ref{tab:gal data}, we assign a fractional systematic error equal to the mean of the fractional systematics from multipipeline objects (see Table~\ref{tab:gal data}). Since the underlying nature of the systematic errors is unknown and represents a potential shift in the central value determination, we account for systematics via the $M_{\text{sys}}$ term in Eq.~\eqref{eq:rhostarLH}, propagating $\delta m_\text{sys}^+$ and $\delta m_\text{sys}^-$ errors separately rather than inaccurately combining with random errors in quadrature.

\begin{figure}[t!]
\centering
\includegraphics[width=0.48\textwidth]{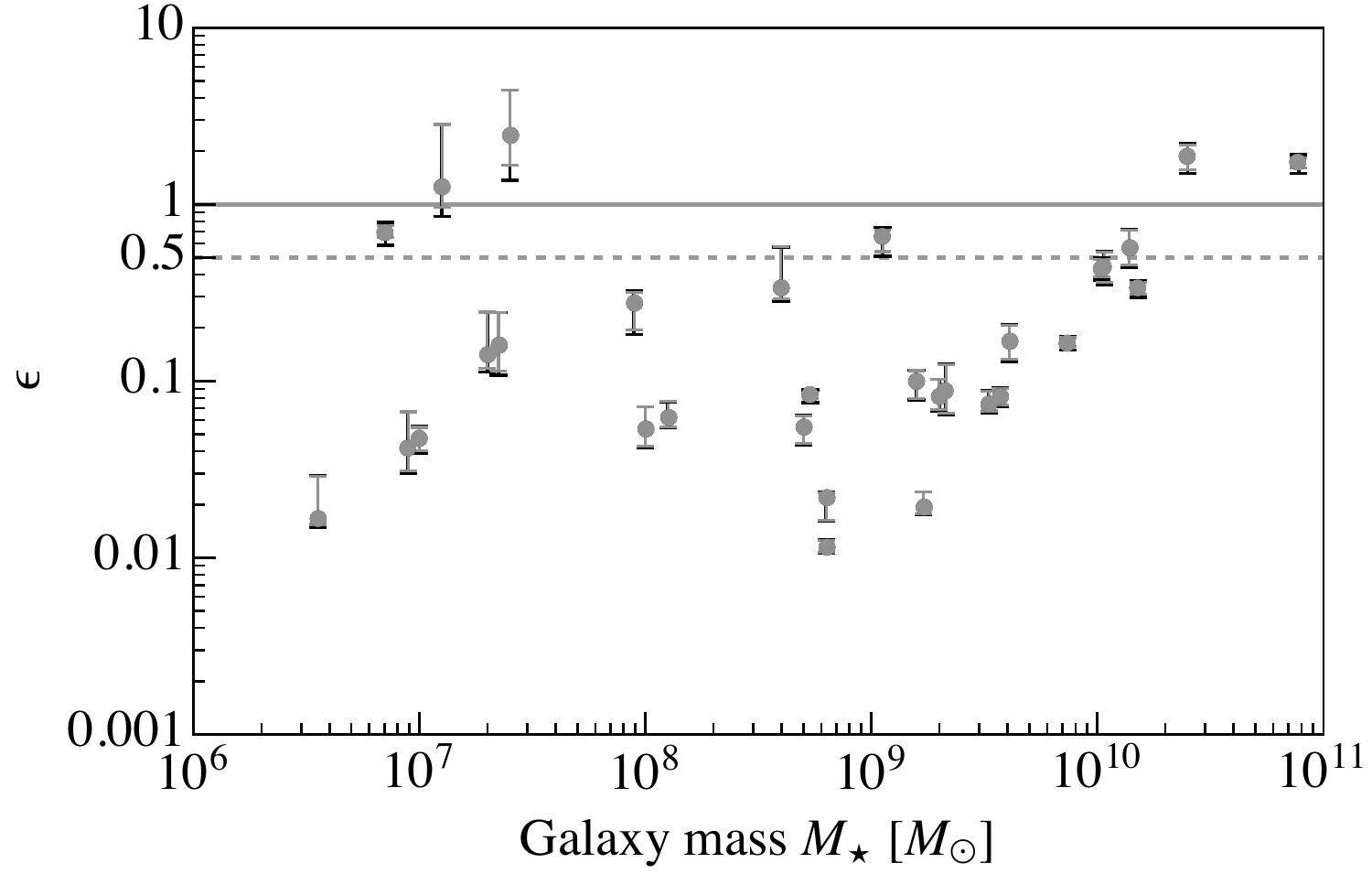}
\caption{Efficiency versus galaxy mass for our sample of 31 galaxies from Table~\ref{tab:gal data}. The solid horizontal line marks a $100\%$ conversion efficiency, and the dashed line indicates $50\%$. Random errors are shown in gray, and random plus sample variance errors are shown in black. For nearly all of the galaxies, the contribution from sample variance is minimal. \label{fig:eff sample var}
}\vskip -0.5cm
\end{figure}

Including the systematic errors separately through the $M_\text{sys}$ term shifts the anchor point $M_\text{halo}$ to $M_\text{halo}\pm(\delta m_\text{sys}^\pm/\epsilon f_b)$ for the error function in Eq.~\eqref{eq:rhostarLH}. Returning to Fig.~\ref{fig:LH}, the shifted masses are indicated by the dashed vertical orange lines, and the two orange curves are the integrands giving the cumulative halo mass density. To fit these two curves on the same linear $y$ axis, they are computed using $10\%$ of the systematic uncertainties; the dark and light orange curves represent the new integrands that incorporate the upper and lower systematic uncertainties, respectively. Clearly, the downward systematic error introduces a much larger cumulative halo mass density. When using the full downward systematic error and random errors [Eq.~\eqref{eq:rhostarLH}], the cumulative halo mass density (and respective cumulative stellar mass density) for a given mass cutoff, is a factor of $\sim\!\!730$ larger than that using the central SED-inferred mass with a standard threshold as in Eq.~\eqref{eq:rhostar}. Since $\rho_\star$ is shifted up by such a factor when including both effects of the systematic error plus random error combination, the inferred required efficiency $\epsilon$ is reduced by orders of magnitude (see position of the observation relative to the sys- curve in Fig.~\ref{fig:MBK sys}). This is the largest effect we find in our analysis of the inferred stellar masses with which to compare the observed galaxies.

In comparison, the upward systematic error only causes a more moderate reduction in the inferred cumulative halo mass density. For our example galaxy, this is attributable to both the effects of the steeply falling density as well as its unequal systematic errors. But, we find that even for galaxies where upper and lower systematic errors are equal in magnitude, the downward systematic error dominates the inferred cumulative halo mass density with which to compare the galaxy. \vskip -0.2cm


\section{\label{sec:results}Results}

Our main result is the finding that the inferred efficiency of star formation is grossly dominated by systematic SED uncertainties on the inferred stellar mass of the JWST detected galaxies and its amplification of the Eddington-like bias. Figure \ref{fig:MBK sys} shows the resulting cumulative comoving stellar mass density curves, for the central value estimates in purple, as well as the deviations including systematic errors in orange. The upper systematic error for $\epsilon$ is acquired by comparing the galaxy point's central value with the sys+ curves, which corresponds to comparison with the \textit{lower} dark orange curves in Fig.~\ref{fig:MBK sys}. We compute analogously for the lower systematic error, which corresponds to the \textit{upper} light orange curves. As shown, the total uncertainty on the inferred efficiency is dominated by systematic errors in the SED-inferred galaxy stellar mass.

Cumulative comoving density plots can only be made for individual galaxies, even for galaxies at the same $z$, given they have differing systematic uncertainties on their SED-inferred stellar mass, which shifts the model curves uniquely in the quantities shown in Fig.~\ref{fig:MBK sys}. To visualize the full sample in a 2D parameter space, we recast the previous comparisons into efficiency–mass and efficiency–$z$ space. For each object, we determine the central value of $\epsilon$ by matching the central galaxy point to the appropriate $\rho_\star$ curve. The same is done for the error bars to propagate the uncertainties described above. Figure~\ref{fig:eff sample var} shows only the values of sample variance in the errors, which is subdominant for most sources. 

\begin{figure*}[t]
\centering
\includegraphics[width=0.48\textwidth]{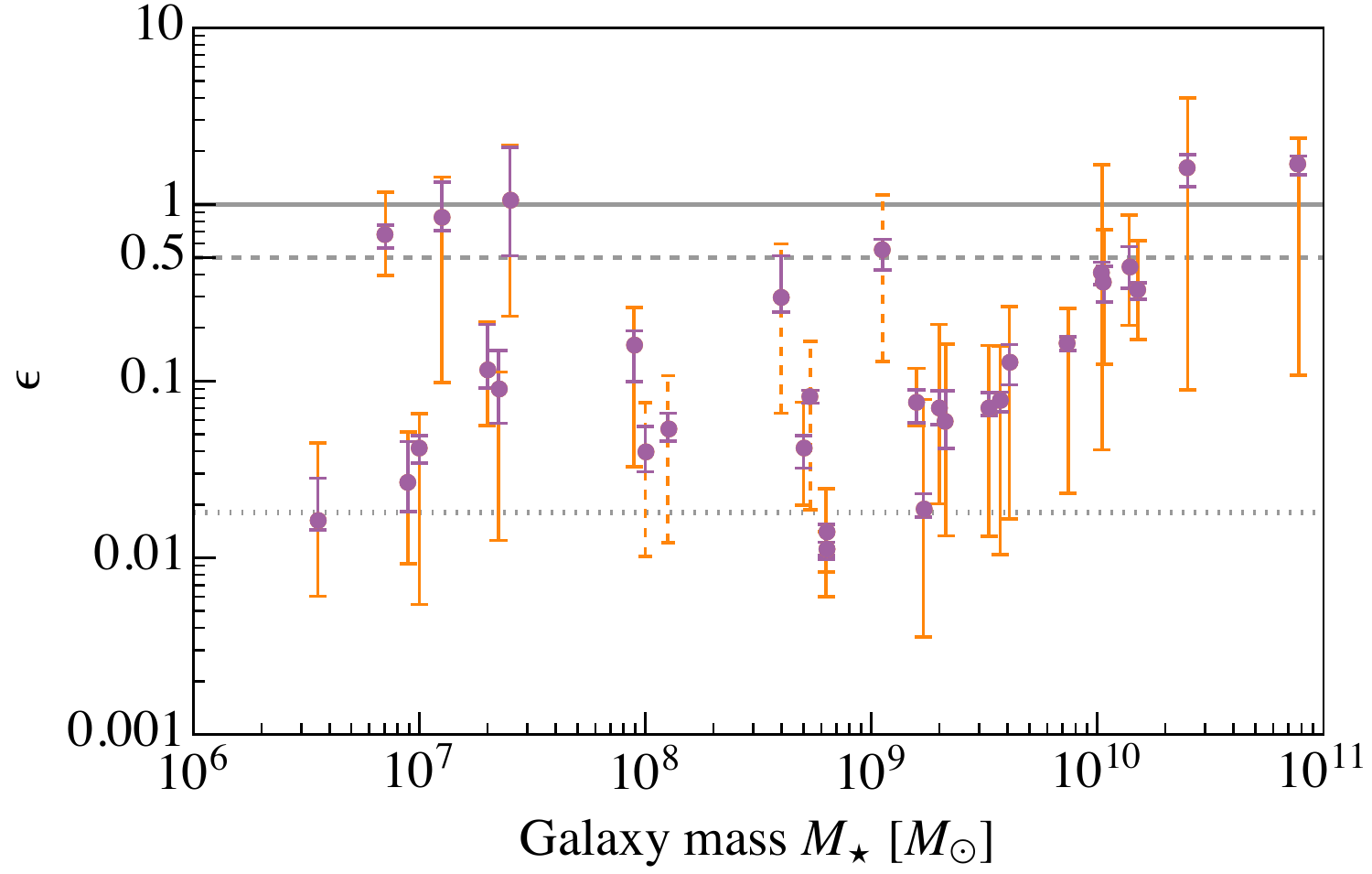}
\includegraphics[width=0.48\textwidth]{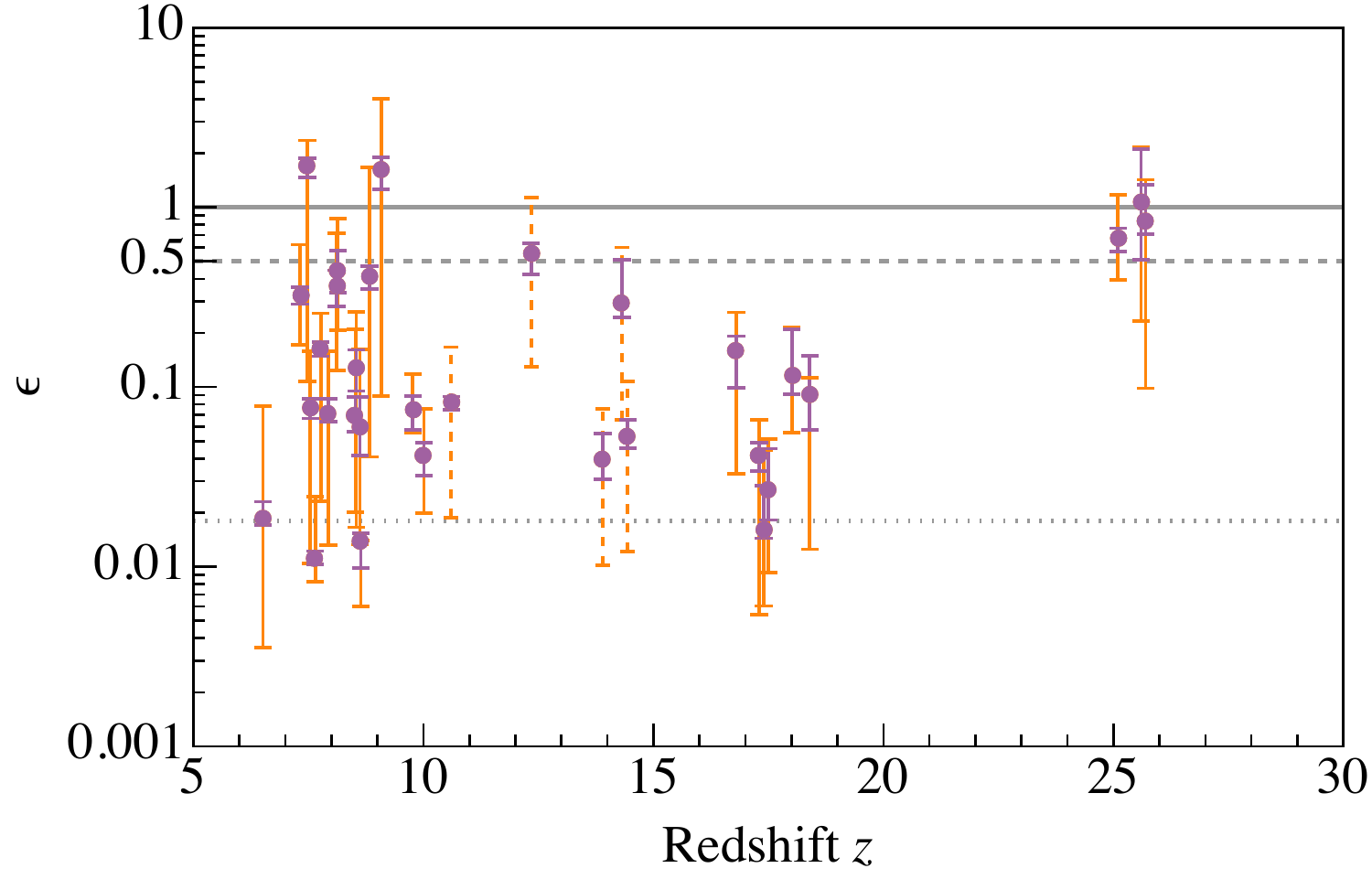}
\caption{Efficiency versus galaxy mass (left) and efficiency versus $z$ (right) for the 31 galaxies from Table~\ref{tab:gal data}. Purple error bars represent random errors plus sample variance, while orange error bars indicate systematic uncertainties. Dashed orange bars correspond to galaxies for which the systematic uncertainty was estimated. As described in the text, we calculate the inferred average $\epsilon_\mathrm{avg} = 0.018 \pm 0.004$, shown as a dotted line. \label{fig:eff sys}  
}\vskip -0.5cm
\end{figure*}

In Fig.~\ref{fig:eff sys}, incorporating the LH correction for random uncertainties, ignoring the $M_\mathrm{sys}$ term in Eq.~\eqref{eq:rhostarLH}, shifts objects toward lower $\epsilon$, with the largest shifts occurring for galaxies that have larger lower-mass random uncertainties. Incorporating SED systematics through the $M_\mathrm{sys}$ further broadens the allowed range shown in Fig.~\ref{fig:eff sys}: even for the most massive objects, the combined uncertainties typically extend below $\epsilon \sim 0.4$. For galaxies analyzed with a single SED pipeline, where systematic errors were estimated rather than directly measured, these are indicated with dashed error bars. The spectroscopically confirmed galaxies from Haro \textit{et al.} \cite{haro_spectroscopic_2023}, where multiple SED pipelines are used, have lower systematic errors than what we see for photometric samples. This suggests spectroscopic confirmation of these JWST galaxies as a potential path to tighten the systematic errors. 

In general, systematic errors are inappropriate to combine with random uncertainties since the nature of the effects of systematic unknowns is nontrivial. However, for purely illustrative purposes, we add our determined systematic errors in quadrature to the random uncertainties to show their relative impact in inferring galaxy formation efficiency. We find the efficiency, $\epsilon$, through a simple $\chi^2$ fit with the combined errors. For the full sample, the conversion efficiency is $\epsilon_\mathrm{avg} = 0.018 \pm 0.004$, with a goodness of fit, $\chi^2_\mathrm{min}/\mathrm{d.\!o.\!f.} = 0.94$ (degrees of freedom). This surprisingly low efficiency is due to the large uncertainties of the highest-efficiency galaxies, with the fit primarily driven by objects with lower efficiencies and smaller errors. Note that the plots in Fig.~\ref{fig:eff sys} have logarithmic axes.\vskip -0.2cm


\section{\label{sec:lum}Luminosity Functions}

Many tests of new physics in galaxy formation at high $z$ now employ the UV luminosity function $\phi_\mathrm{UV}$. Such tests require a mapping between the halo mass and luminosity of the galaxies \cite{Vale:2004yt,Cooray:2005mm,Wechsler:2018pic,Blamart:2025szc}. Though it is beyond the scope of this paper to apply our presented methods to observations of $\phi_\mathrm{UV}$, we develop the formalism that includes random and systematic errors to the theory and modeling of $\phi_\mathrm{UV}$, and leave future tests including these effects to future work. One can implement bias in the probabilistic description of the  $\phi_\mathrm{UV}$ from JWST as the following:
\begin{widetext}
\begin{align}
    \phi_\mathrm{UV} \equiv \frac{dn}{dM_\mathrm{UV}} &= \int dM \frac{dn}{dM} P_\mathrm{c}(M_\mathrm{UV}|M)
    + \int dM_\mathrm{s}\frac{dn_\mathrm{s}}{dM_\mathrm{s}} P_\mathrm{s}(M_\mathrm{UV}|M_\mathrm{s})\, , 
\end{align}
where $dn/dM$ is the full halo mass function, and $dn_s/dM_s$ is the subhalo mass function, and $P_\mathrm{c}(M_\mathrm{UV}|M)$ ($P_\mathrm{s}(M_\mathrm{UV}|M_\mathrm{s})$) is the probability that a halo (subhalo) of mass $M$ ($M_\mathrm{s}$) hosts a central (satellite) galaxy of magnitude $M_\mathrm{UV}$. This is often adopted to be Gaussian in $\ln M_\mathrm{UV}$ centered around $M_\mathrm{UV}^\mathrm{avg}(M)$. The binned luminosity function of galaxies $\Phi_{\mathrm{UV},i}$ with magnitude between $M_{\mathrm{UV},i}$ and $M_{\mathrm{UV},i+1}$ is then 
\begin{align}
    \Phi_{\mathrm{UV},i} &= \int_{M_{\mathrm{UV},i}}^{M_{\mathrm{UV},i+1}} dM_\mathrm{UV}\,\phi_\mathrm{UV} \nonumber \\ 
    &= \int_{M_{\mathrm{UV},i}}^{M_{\mathrm{UV},i+1}} dM_\mathrm{UV}\, \left[ \int dM \frac{dn}{dM} P_\mathrm{c}(M_\mathrm{UV}+M_{\mathrm{UV},i}^\mathrm{sys}|M)
    + \int dM_\mathrm{s}\frac{dn_\mathrm{s}}{dM_\mathrm{s}} P_\mathrm{s}(M_\mathrm{UV}+M_{\mathrm{UV},i}^\mathrm{sys}|M_\mathrm{s}) \right]\,.
\end{align}
\end{widetext}
Here, the Gaussian accounts for the Eddington-like bias, and can incorporate both random and systematic errors. For the case of these massive high-$z$ JWST galaxies, one may want to consider the case of one observed galaxy per halo, i.e., the halo occupation fraction is unity, and therefore, the subhalo term can be ignored. 

Adopting a halo occupation fraction of unity, we get,
\begin{align}
    \Phi_{\mathrm{UV},i} &= \int dM \frac{dn}{dM}\int_{M_{\mathrm{UV},i}}^{M_{\mathrm{UV},i+1}}\!\!\!\!\! dM_\mathrm{UV}\, P_\mathrm{c}(M_\mathrm{UV}+M_{\mathrm{UV},i}^\mathrm{sys}|M) \nonumber \\
    &= \int \frac{dM}{M} \frac{dn}{d\ln M} \frac{1}{2}\left[ \mathrm{erfc}\left(y_i \right)-\mathrm{erfc}\left(y_{i+1} \right)\right]\,,
\end{align}
where $y_i$ is defined as
\begin{align}
    y_i(M) \equiv \frac{\ln M_{\mathrm{UV},i} - \ln M_\mathrm{UV}^\mathrm{avg}(M) - \ln M_{\mathrm{UV},i}^\mathrm{sys}}{\sqrt{2 \sigma_{\ln M_{\mathrm{UV},i}}}}\, .
\end{align}
Here $\sigma_{\ln M_\mathrm{UV,i}}^2 \equiv \sigma_{\ln M_\mathrm{rand,UV,i}}^2 +\sigma_{\ln M_\mathrm{int,UV,i}}^2$ includes the intrinsic scatter in the conditional luminosity function to halo mass relationship, $\sigma_{\ln M_\mathrm{int,UV}}$, and the statistical measurement error, $\sigma_{\ln M_\mathrm{rand,UV}}$. If one includes subhalo galaxy contributions, there would be a corresponding second pair of binned complementary error functions. \vskip -0.2cm

\section{\label{sec:conc} Discussion and Conclusions}

In this work, we have tested the compatibility of the most massive high-$z$ JWST galaxies with $\Lambda$CDM using abundance matching in the halo model framework, explicitly accounting for the major sources of this framework's uncertainty, both random and systematic, in order to test whether structure formation models beyond $\Lambda$CDM are required. We examined the model dependence introduced by different HMFs and found that while discrepancies are noticeable at lower masses, the high-$z$ calibrated HMFs agree closely in the high-mass regime relevant for these galaxies. We also incorporated sample variance effects arising from LSS, finding that it is significant, but not dominant. Most importantly, we analyzed the Eddington-like bias presented by random errors, SED-driven systematic errors, and their combination. We found that the systematic errors introduced by SED-fitting methods to determine the galaxies' stellar mass are the dominant errors in the determination of the inferred baryon-to-stellar mass efficiency. This last effect alleviates much of the apparent tension between the most massive JWST galaxies and $\Lambda$CDM abundance expectations. Within our framework, all objects are consistent with $\epsilon \lesssim 0.4$ once all uncertainties are included. When systematic and random uncertainties are combined in quadrature for a purely illustrative $\chi^2$ fit, the best-fit global efficiency is $\epsilon_\mathrm{avg} = 0.018 \pm 0.004$. Under this set of assumptions, one may cautiously state that the data may be consistent with modest baryon-to-stellar-mass conversion efficiencies under $\Lambda$CDM. However, some galaxies still require high efficiencies, e.g., $\epsilon > 0.4$ for midis-z25-2 \cite{Perez-Gonzalez:2025bqr}.

As observational constraints improve, it remains possible that some high-$z$ galaxies will be confirmed to form stars at substantially higher efficiencies. Note that large samples of galaxies at $z>4$ in the CEERS Survey indicate that the efficiencies of conversion of baryons to stars increases with redshift, potentially consistent with that expected at high $z$ \cite{chworowsky_evidence_2023}. Moreover, at $z > 10$, $\epsilon$ could potentially approach unity \cite{Dekel:2023ddd}, which is also seen in high resolution, high-$z$ simulations and semianalytic models \cite{Keller:2022mnb,McCaffrey_NoTension_2023,Robertson:2022gdk}. 

Two caveats are worth noting: First, we have not applied volume incompleteness corrections within the $z$ bins. Because incompleteness primarily affects the contribution from less massive systems, we compare cumulative stellar mass densities above a threshold. This is a conservative assumption, and should have a limited impact on the inferred densities. Second, slit/aperture losses and surface brightness selection may bias photometry and thus SED masses, particularly for compact, high-$z$ sources, some of which may have an active galactic nucleus component~\cite{Calvi:2014zqa,Kocevski:2025tft}.

Future statistical methods may jointly constrain $\epsilon$ and nuisance parameters associated with SED systematic errors and selection effects. Here, we have adopted using medians to remain agnostic to the origin of the systematic effects of SED-derived masses. On the observational side, expanding the set of spectroscopically confirmed high-$z$ sources and cross-comparing multiple SED pipelines will provide a clearer picture. The large systematic uncertainties are already reduced for spectroscopically confirmed galaxies, as seen in the Haro \textit{et al.}\ \cite{haro_spectroscopic_2023} samples in Table~\ref{tab:gal data}. As new galaxies are detected, and currently detected ones are followed up, the picture of the masses and ages of the galaxies will be further refined to test new galaxy formation and dark matter structure formation models \cite{2024MNRAS.530..966G, 2024ApJ...963..128B, 2023ApJ...946L..16P, 2024ApJ...969L..13W, 2024MNRAS.529..855W, 2024ApJ...969L...2F, Andika:2024jto, Fujimoto:2023orx, 2024MNRAS.527.5004M, 2023MNRAS.520.4554D, 2023MNRAS.519.5076B, Heintz:2025ijn}. 

Several analyses are testing early galaxy formation through the UV luminosity function. We show that the observed binned UV luminosity is subject to similar requirements of proper inclusion of random and systematic errors. The qualitative and quantitative difference is that $\Phi_\mathrm{UV}$ is a binned luminosity density instead of a cumulative stellar mass density. That is, the kernel mapping from the HMF to the observable is the probabilistic conditional luminosity function instead of the abundance matching of the fractional of matter density as baryons forming stars. 

Many new physics models have been proposed to accommodate the observed JWST galaxies, including dynamical dark energy \cite{Menci:2022wia}, enhanced clustering from massive primordial black holes \cite{Liu:2022bvr}, non-Gaussianities  \cite{Biagetti:2022ode}, axion miniclusters \cite{Hutsi:2022fzw}, axion quark nuggets \cite{Zhitnitsky:2023znn},  domain walls  \cite{Guo:2023hyp}, or ultralight axion dark matter fragmentation \cite{Bird:2023pkr}, or even an early negative cosmological constant \cite{Adil:2023ara}. Overall, our work shows that, once statistical and systematic uncertainties are properly accounted for, the observed early galaxies from JWST are largely reconciled with $\Lambda$CDM expectations. We have provided an openly available framework linking the galaxy detections' inferred stellar mass densities to the underlying HMF and LSS of the primordial matter power spectrum arising from the early Universe \cite{JWSTEG} that accounts for all the sources of uncertainty described in this paper. Our framework can test whether new JWST detections of massive galaxies, or current high-$z$ galaxies with reduced uncertainties, will reveal new insights into the earliest stages of structure formation arising from inflation or dark-matter-induced dynamics.\vskip -0.2cm


\section*{Acknowledgments}

\small{We acknowledge useful conversations with Mike Boylan-Kolchin, Mike Cooper, Francis-Yan Cyr-Racine, Andrew Evans, Helena Garc\'ia Escudero, Ryan Keeley, Toni Makela, Michael Ryan, Tyler Smith, and Cannon Vogel. 
This material is based upon work by J. R. K. supported by the GAANN Fellowship from the US Department of Education under Grant No.\ P200A240014, and the National Science Foundation Graduate Research Fellowship (NSF) Program under Grant No.\ DGE-2235784, and upon work by K. N. A. supported by NSF Theoretical Physics Program Grant No.\ PHY-2210283. 
 
Any opinions, findings, and conclusions or recommendations expressed in this material are those of the author(s) and do not necessarily reflect the views of the NSF.}\vskip -0.5cm


\section*{Data Availability}

\small{The data that support the findings of this article are openly available~\cite{JWSTEG}.}


\bibliographystyle{apsrev4-2}
\bibliography{JWST}

\begin{table*}[t]
\centering

\caption{
Summary of JWST galaxy data used in this analysis from Refs.~\cite{Labbe:2022ahb, haro_spectroscopic_2023, Bunker:2023lzn, castellano_jwst_2024, carniani_spectroscopic_2024, Naidu:2025xfo, Perez-Gonzalez:2025bqr}. Galaxy IDs correspond to those provided in the original studies. Errors shown for redshift and galaxy mass are statistical uncertainties. Systematic errors for galaxy masses are those provided by the authors, or as noted below. Inferred efficiencies have statistical, including sample variance, and systematic errors shown in that order. 
\label{tab:gal data}}

    \renewcommand{\arraystretch}{1.4} 
    \begin{tabular}{@{\hspace{6pt}}c@{\hspace{6pt}}|@{\hspace{6pt}}c@{\hspace{6pt}}|@{\hspace{6pt}}c@{\hspace{6pt}}|@{\hspace{6pt}}c@{\hspace{6pt}}|@{\hspace{6pt}}c@{\hspace{6pt}}|@{\hspace{6pt}}c@{\hspace{6pt}}|@{\hspace{6pt}}c@{\hspace{6pt}}
    }
      \toprule \midrule[0.3pt]
      Reference & galaxy ID & redshift $z$ & galaxy mass & systematic error & FOV area & Inferred efficiency\\
       & & & $\left(\log_{10}{M_\star/M_\odot}\right)$ & $\left(\log_{10}{M_\star/M_\odot}\right)$ & $($arcmin$^2$) & $\epsilon\substack{\,\mathrm{+rand}\,\mathrm{+sys} \\ \,\mathrm{+rand}\,\mathrm{+sys}}$\\
      \midrule 
      Labbé \textit{et al.}\ & 2859 & $8.11 \substack{+0.49 \\ -1.49}$ & $10.03 \substack{+0.24 \\ -0.27}$ & $(+0.46,-0.75)$ & 38 & $0.362 \substack{+0.084 \,+0.357 \\ -0.081 \,-0.238 }$\\
      Labbé \textit{et al.}\ & 7274 & $7.77 \substack{+0.05 \\ -0.06}$ & $9.87 \substack{+0.09 \\ -0.06}$ & $(+0.30,-1.36)$ & 38 & $0.162 \substack{+0.015 \,+0.095 \\ -0.014 \,-0.139 }$\\
      Labbé \textit{et al.}\ & 11184 & $7.32 \substack{+0.28 \\ -0.35}$ & $10.18 \substack{+0.10 \\ -0.10}$ & $(+0.42,-0.43)$ & 38 & $0.325 \substack{+0.034 \,+0.295 \\ -0.035 \,-0.153 }$\\
      Labbé \textit{et al.}\ & 13050 & $8.14 \substack{+0.45 \\ -1.71}$ & $10.14 \substack{+0.29 \\ -0.30}$ & $(+0.45,-0.54)$ & 38 & $0.446 \substack{+0.130 \,+0.422 \\ -0.111 \,-0.240 }$\\
      Labbé \textit{et al.}\ & 14924 & $8.83 \substack{+0.17 \\ -0.09}$ & $10.02 \substack{+0.16 \\ -0.14}$ & $(+0.90,-1.63)$ & 38 & $0.408 \substack{+0.062 \,+1.260 \\ -0.058 \,-0.368 }$\\
      Labbé \textit{et al.}\ & 16624 & $8.52 \substack{+0.19 \\ -0.22}$ & $9.30 \substack{+0.27 \\ -0.24}$ & $(+0.72,-0.87)$ & 38 & $0.070 \substack{+0.018 \,+0.139 \\ -0.014 \,-0.050 }$\\
      Labbé \textit{et al.}\ & 21834 & $8.54 \substack{+0.32 \\ -0.51}$ & $9.61 \substack{+0.26 \\ -0.32}$ & $(+0.49,-1.50)$ & 38 & $0.128 \substack{+0.033 \,+0.135 \\ -0.033 \,-0.111 }$\\
      Labbé \textit{et al.}\ & 25666 & $7.93 \substack{+0.09 \\ -0.16}$ & $9.52 \substack{+0.23 \\ -0.10}$ & $(+0.52,-1.17)$ & 38 & $0.071 \substack{+0.015 \,+0.088 \\ -0.007 \,-0.058 }$\\
      Labbé \textit{et al.}\ & 28984 & $7.54 \substack{+0.08 \\ -0.14}$ & $9.57 \substack{+0.13 \\ -0.15}$ & $(+0.47,-1.42)$ & 38 & $0.077 \substack{+0.009 \,+0.081 \\ -0.010 \,-0.067 }$\\
      Labbé \textit{et al.}\ & 35300 & $9.08 \substack{+0.31 \\ -0.38}$ & $10.40 \substack{+0.19 \\ -0.23}$ & $(+0.60,-2.11)$ & 38 & $1.605 \substack{+0.301 \,+2.402 \\ -0.344 \,-1.516 }$\\
      Labbé \textit{et al.}\ & 37888 & $6.51 \substack{+1.42 \\ -0.28}$ & $9.23 \substack{+0.25 \\ -0.10}$ & $(+0.92,-1.17)$ & 38 & $0.019 \substack{+0.004 \,+0.060 \\ -0.002 \,-0.015 }$\\
      Labbé \textit{et al.}\ & 38094 & $7.48 \substack{+0.04 \\ -0.04}$ & $10.89 \substack{+0.09 \\ -0.08}$ & $(+0.22,-1.99)$ & 38 & $1.683 \substack{+0.193 \,+0.674 \\ -0.216 \,-1.575 }$\\
      Labbé \textit{et al.}\ & 39575 & $8.62 \substack{+0.34 \\ -0.57}$ & $9.33 \substack{+0.43 \\ -0.39}$ & $(+0.69,-1.11)$ & 38 & $0.060 \substack{+0.029 \,+0.102 \\ -0.018 \,-0.046 }$\\
      Haro \textit{et al.}\ & 80041 & $10.01 \substack{+0.14 \\ -0.19}$ & $8.7 \substack{+0.2 \\ -0.3}$ & $(+0.4,-0.5)^*$ & 66.4$^\ddagger$ & $0.041 \substack{+0.008 \,+0.034 \\ -0.009 \,-0.022 }$\\
      Haro \textit{et al.}\ & 80026 & $9.77 \substack{+0.37 \\ -0.29}$ & $9.2 \substack{+0.2 \\ -0.3}$ & $(+0.3,-0.2)^*$ & 66.4$^\ddagger$ & $0.075 \substack{+0.014 \,+0.043 \\ -0.017 \,-0.019 }$\\
      Haro \textit{et al.}\ & 80083 & $8.638 \substack{+0.001 \\ -0.001}$ & $8.8 \substack{+0.1 \\ -0.4}$ & $(+0.0,-0.6)^*$ & 66.4$^\ddagger$ & $0.014 \substack{+0.001 \,+0.000 \\ -0.004 \,-0.008 }$\\
      Haro \textit{et al.}\ & 80025 & $7.651 \substack{+0.001 \\ -0.001}$ & $8.8 \substack{+0.1 \\ -0.1}$ & $(+0.5,-0.2)^*$ & 66.4$^\ddagger$ & $0.011 \substack{+0.001 \,+0.013 \\ -0.001 \,-0.003 }$\\
      Bunker \textit{et al.}\ & GN-z11 & $10.6034 \substack{+0.0013 \\ -0.0013}$ & $8.73 \substack{+0.06 \\ -0.06}$ & $(+0.46,-1.01)^\dagger$ & 22.98$^\ddagger$ & $0.082 \substack{+0.006 \,+0.085 \\ -0.007 \,-0.063 }$\\
      Castellano \textit{et al.}\ & GLASS-z12 & $12.342 \substack{+0.009 \\ -0.009}$ & $9.05 \substack{+0.10 \\ -0.25}$ & $(+0.48,-1.05)^\dagger$ & 9.76$^\ddagger$ & $0.559 \substack{+0.075 \,+0.574 \\ -0.135 \,-0.429 }$\\
      Carniani \textit{et al.}\ & 183348 & $14.32 \substack{+0.08 \\ -0.20}$ & $8.6 \substack{+0.7 \\ -0.2}$ & $(+0.4,-1.0)^\dagger$ & 58 & $0.295 \substack{+0.217 \,+0.302 \\ -0.051 \,-0.230 }$\\
      Carniani \textit{et al.}\ & 18044 & $13.90 \substack{+0.17 \\ -0.17}$ & $8.0 \substack{+0.4 \\ -0.3}$ & $(+0.4,-0.9)^\dagger$ & 58 & $0.040 \substack{+0.015 \,+0.036 \\ -0.009 \,-0.030 }$\\
      Naidu \textit{et al.}\ & MoM-z14 & $14.44 \substack{+0.02 \\ -0.02}$ & $8.1 \substack{+0.3 \\ -0.2}$ & $(+0.4,-0.9)^\dagger$ & 350 & $0.053 \substack{+0.013 \,+0.054 \\ -0.007 \,-0.041 }$\\
      Pérez-González \textit{et al.}\ & midis-z17-1 & $18.4 \substack{+1.5 \\ -1.5}$ & $7.35 \substack{+0.55 \\ -0.45}$ & $(+0.15,-1.45)^*$ & 17.6 & $0.091 \substack{+0.058 \,+0.022 \\ -0.033 \,-0.078 }$\\
      Pérez-González \textit{et al.}\ & midis-z17-2 & $17.4 \substack{+1.4 \\ -1.1}$ & $6.55 \substack{+0.70 \\ -0.10}$ & $(+0.65,-0.65)^*$ & 17.6 & $0.016 \substack{+0.012 \,+0.028 \\ -0.002 \,-0.010 }$\\
      Pérez-González \textit{et al.}\ & midis-z17-3 & $17.3 \substack{+1.1 \\ -0.2}$ & $7.00 \substack{+0.20 \\ -0.20}$ & $(+0.30,-1.40)^*$ & 17.6 & $0.041 \substack{+0.008 \,+0.024 \\ -0.007 \,-0.036 }$\\
      Pérez-González \textit{et al.}\ & midis-z17-4 & $18.0 \substack{+1.7 \\ -1.5}$ & $7.30 \substack{+0.70 \\ -0.25}$ & $(+0.40,-0.50)^*$ & 17.6 & $0.117 \substack{+0.092 \,+0.098 \\ -0.025 \,-0.061 }$\\
      Pérez-González \textit{et al.}\ & midis-z17-5 & $17.5 \substack{+2.3 \\ -2.4}$ & $6.95 \substack{+0.60 \\ -0.40}$ & $(+0.45,-0.75)^*$ & 17.6 & $0.027 \substack{+0.019 \,+0.025 \\ -0.008 \,-0.017 }$\\
      Pérez-González \textit{et al.}\ & midis-z17-6 & $16.8 \substack{+2.4 \\ -0.8}$ & $7.95 \substack{+0.20 \\ -0.45}$ & $(+0.35,-1.15)^*$ & 17.6 & $0.158 \substack{+0.034 \,+0.102 \\ -0.059 \,-0.125 }$\\
      Pérez-González \textit{et al.}\ & midis-z25-1 & $25.7 \substack{+1.8 \\ -1.4}$ & $7.10 \substack{+1.05 \\ -0.35}$ & $(+0.40,-1.50)^*$ & 17.6 & $0.849 \substack{+0.489 \,+0.577 \\ -0.140 \,-0.751 }$\\
      Pérez-González \textit{et al.}\ & midis-z25-2 & $25.1 \substack{+1.5 \\ -1.0}$ & $6.85 \substack{+0.10 \\ -0.10}$ & $(+0.35,-0.35)^*$ & 17.6 & $0.676 \substack{+0.089 \,+0.494 \\ -0.111 \,-0.282 }$\\
      Pérez-González \textit{et al.}\ & midis-z25-3 & $25.6 \substack{+1.5 \\ -1.6}$ & $7.40 \substack{+0.75 \\ -0.55}$ & $(+0.50,-1.10)^*$ & 17.6 & $1.060 \substack{+1.047 \,+1.102 \\ -0.548 \,-0.827 }$\\
      \midrule[0.3pt]\bottomrule 
    \end{tabular}

\vspace{6pt}

\parbox{0.95\linewidth}{\footnotesize \raggedright

\textit{Notes:} \\
\hspace{1.5em}* Systematic uncertainties reflect the extrema relative to the median stellar masses derived from different SED-fitting models.\\
\hspace{1.5em}$\dagger$ For galaxies with only a single SED fit, the mean fractional systematic error from multi-pipeline objects is applied.\\
\hspace{1.5em}$\ddagger$ The entire field of view (FOV) of field~\cite{Conselice:2024yls}.
}
\end{table*}

\end{document}